\documentclass[11pt,twoside]{article}
\usepackage{./asp2014}

\aspSuppressVolSlug
\resetcounters

\begin{document}

\title{The Challenges of Observing, Calibrating, and Modeling Stellar Spectral Energy Distributions}
\author{Carlos Allende Prieto,$^1$
\affil{$^1$Instituto de Astrof\'{\i}sica de Canarias, Via L\'actea s/n, 38205 La Laguna, Tenerife, Spain; \email{callende@iac.es}}}

% This section is for ADS Processing.  There must be one line per author.
\paperauthor{Carlos Allende Prieto}{callende@iac.es}{0000-0002-0084-572X}{Instituto de Astrof\'{\i}sica de Canarias}{}{La Laguna}{Tenerife}{38205}{Spain}

\begin{abstract}
While optical and quantum efficiency are on the rise, and spectrographs becoming massively multiplexed, measuring  spectral energy distributions of astronomical sources with accuracy remains a challenge. In addition to atmospheric refraction, extinction, variability, and limited apertures of instrument entrance slits and optical fibers, accurate calibration is hampered by issues such as a limited choice of reliable standard stars. Modeling stellar spectral energy distributions has seen good progress, but some weaknesses survive, especially for late-type stars. This article provides an overview of these matters and discusses observation, calibration, and modeling strategies for future large spectroscopic surveys.
\end{abstract}

\section{The information in SEDs}

Spectral energy distributions (SEDs) are shaped, to first order, by the radiative flux at the bottom of stellar photospheres, and therefore they are a good thermometer. The fact that the continuum is formed in deep photospheric layers implies high densities and conditions where local thermodynamical equilibrium (LTE) is more likely to prevail. Overall, we can expect SEDs to be a robust prediction of model atmospheres.
SEDs are also shaped by line and continuous opacity. This is second order compared to the radiative flux (i.e. the stellar effective temperature). but it means they are useful to constrain surface gravity and chemical abundances. 

Despite deeper into the photosphere convection gets stronger, colors and broad-band SEDs are not seriously affected by 3D effects. Chiavassa et al. (2018) have recently made a detailed study on this matter, and concluded that hydrodynamical models do not induce changes in low resolution SEDs much larger than about 5 \%. 

Fig. \ref{f1} illustrates the impact of some of the most relevant elements on the solar SED. Carbon (CN and CH) provides strong line absorption in the blue.  Magnesium contributes some strong lines (atomic Mg and MgH) and substantial continuum opacity bluewards of 250 nm; it also provides free electrons that enhance the H$^{-}$ opacity. Aluminum imprints few strong lines and gives continuum opacity in the far UV. Silicon contributes strong lines, in addition to bringing significant free electrons that enhance the  H$^{-}$ opacity. Calcium provides strong neutral and singly-ionized lines. Finally, iron provides free electrons, its own (Fe I) continuum opacity bluewards of 400 nm, and a myriad of atomic lines all over the optical. An in-depth study of each contributor is warranted, 
given that updated atomic and molecular data are
available.

\articlefigure[width=.9\textwidth]{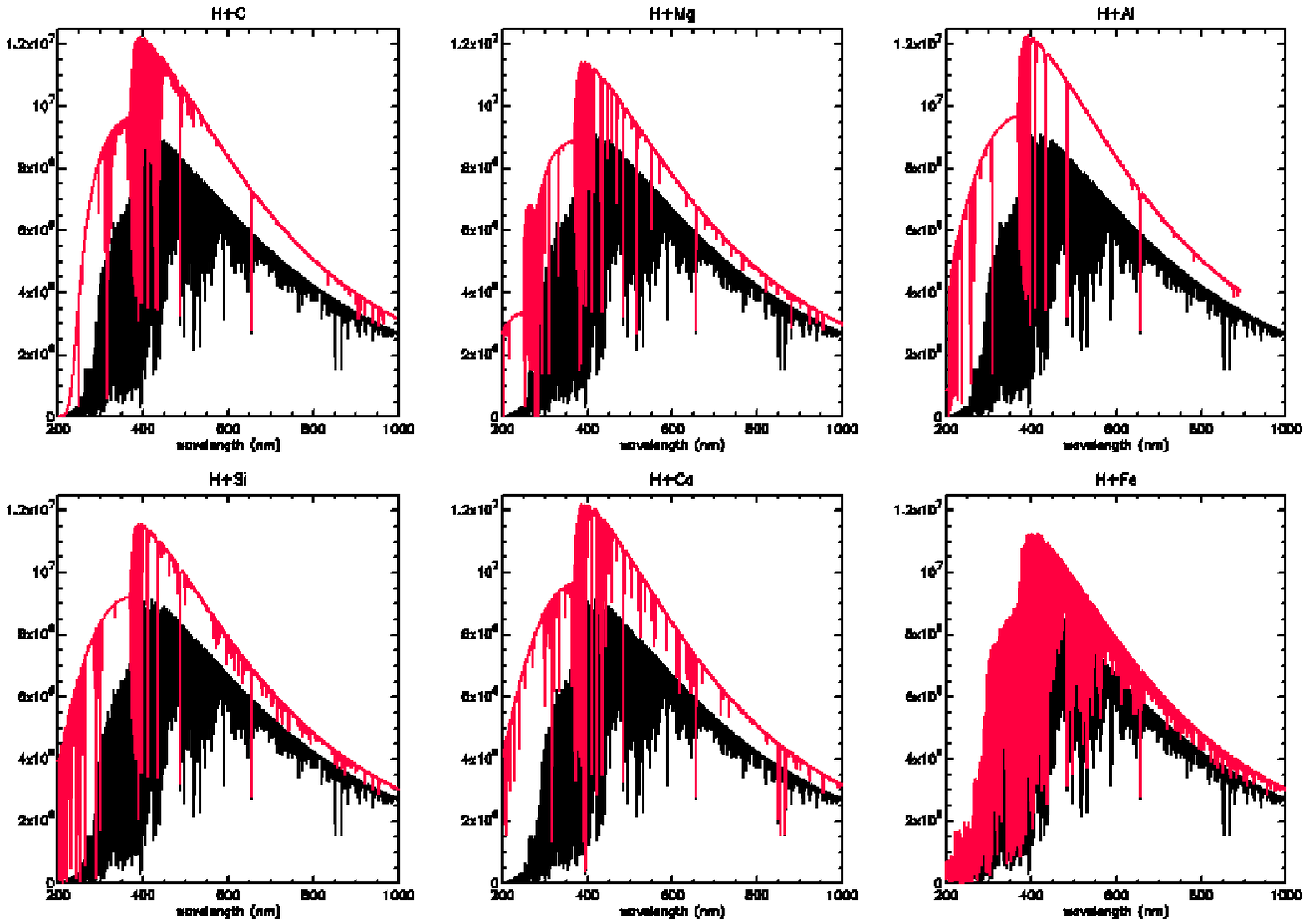}{f1}{Impact of various elements on the solar SED.}
% It is possible to reduce the size of a figure among other changes (see the instructions).  Here is an example:
% \articlefigure[width=.5\textwidth]{example.eps}{ex_fig1_reduced}{Welcome to 1953 a little smaller.}

%\articlefiguretwo{f1.eps}{f1.eps}{ex_fig2}{Now there are two of them.  \emph{Left:} An image from long ago.  \emph{Right:} The same exact thing.}
% There is a figure command allowing for three figures:
% \articlefigurethree{example.eps}{example.eps}{example.eps}{ex_fig1_triple}{Now there are three of them.}

With sufficiently high resolution, SEDs contain rich information about stars. They allow us to quantify the strength of the H lines, which provide strong constraints on the stellar effective temperature. Molecular absorption such as  
CH or MgH bands depend strongly on surface gravity, and so does the Balmer jump at 370 nm. Spectral lines, in general, provide valuable information on (micro)turbulence and chemistry. When the SED measurements 
span a wide wavelength range, and there is an estimate of the stellar angular diameter, they directly provide a determination of effective temperature. The angular diameter may be derived 
from interferometry for targets that are large enough and near enough. They can also be derived 
from the ratio of the fluxes observed and those predicted by models. This requires knowledge of
the stellar atmospheric parameters, but as shown in previous studies, the dependency on surface
gravity and metallicity is relatively small (del Burgo et al. 2010; Allende Prieto \& del Burgo  2016)

\section{Observations}

SED measurements require good photometry plus good spectroscopy. This involves control of the instrumental response, photometric conditions (or avoiding entirely the atmosphere from space), and wide slits and, in the case of ground observations, carrying the measurements at low airmass and at the parallactic angle to avoid/minimize differential atmospheric refraction. 

As shown in the literature, and in particular the papers published by R. Bohlin (e.g. Bohlin 1996), ideal conditions can be achieved from space, with very wide (or no) slits and careful calibration procedures. One can also do a very good job from the ground using a wide (and long) slit, excellent observing conditions, and nearby reference standards. In an unpublished experiment, I tracked one of the most studied planet-hosting stars, HD 209458, with the Intermediate Dispersion Spectrograph (IDS), on the 2.5-m INT in La Palma. The observations alternated between HD 209458 and BD $+17$ 4708, the SDSS standard, which enjoys STIS spectrophotometry (Bohlin \& Gilliland 2004). 
By tracking the variations with airmass of
the spectrum of the standard, it was possible to calibrate the spectrophotometry of HD 290458 to 2-3 \%. This precision level was easy to check since there is STIS spectrophotometry available for HD 209458.

Modern instruments for medium resolution spectroscopy are not like the IDS. The tendency goes towards larger fields of view and massive multiplexing: hundreds to thousands of objects observed simultaneously, usually employing optical fibers. Fibers are tougher to handle for spectrophotometry than slits. The prime example is the Sloan Digital Sky Survey (SDSS), which in addition of 5-band imaging for 1/3 of the sky, has obtained fiber spectroscopy for millions of targets, mainly quasars, galaxies, and stars. The SDSS telescope (Gunn et al. 2006) offers a 7-degree field of view and the SDSS/BOSS spectrographs (2 twin spectrographs, each with two cameras; Smee et al. 2013) are fed by 1000 fibers (640 fibers between 1999 and 2008, and 1000 thereafter) spreaded over the focal plane. 

The high multiplexing of the new generation of instruments provides an opportunity for mapping sky (absorption and emission) variations in detail, but lacking an atmospheric differential-refraction corrector (ADC) and the use of relatively small, 1-2 arc seconds in diameter, fibers, compromises the accuracy of the spectrophotometric calibration. Despite the large field-of-view, finding standard stars for flux calibration is a challenge. This was circumvented in SDSS by {\it creating} standards: choosing stars with smooth, easy to model, SEDs, charactering these stars from their (uncalibrated) spectra, and then using them to calibrate the rest of the stars observed in a given field. Having dozens of such {\it standard} stars all over the field makes it possible to understand the distortions of the instrument's response over the field and along the spectrograph's slit. 

\section{Standard stars}

SDSS, focused mainly in high Galactic latitude observations, chosen to use F-type halo turn-off stars as {\it standards}. These stars are abundant enough, and can be easily identified on the basis of their colors, using BD $+17$ 4708 as a reference. These stars are metal deficient, with a mean metallicity of [Fe/H] $\simeq -1.5$, exhibit a relatively simple continuum dominated by atomic H and H$^{-}$ opacity, and a modest number of lines, mainly hydrogen lines and some strong iron and calcium ones. 

As extensively described in the literature (Bohlin 2000; Allende Prieto et al. 2009), DA white dwarfs are even better calibrators, but far too scarce and faint to be useful for large-scale surveys. Another possibility are using A-type stars (Allende Prieto \& del Burgo 2016). The optical spectrum of A-type stars is dominated by atomic H continuum and line opacity, with some Fe II lines visible in the near UV. They are more abundant than white dwarfs, especially on the Galactic plane, but they are harder to model. 

In preparation for the Dark Energy Spectroscopic Instrument (DESI Collaboration  2016a, b), we have recently made estimates of the densities of the {\it standard} stellar classes mentioned above. DESI is bound to replace the SDSS with a larger telescope (4-m Mayall at Kitt Peak), yet a 7-degree diameter field-of-view, 5000 robotically-positioned fibers, and higher spectral resolution than SDSS. In the typical (high Galactic latitude) DESI field, at magnitudes in the range $16<g<19$, there are about 40 F-type turn-off stars, about 10 A-type stars, and about 2 DA white dwarfs, per square degree. We can hopefully use the F- and A-type stars for calibration, reserving the white dwarfs to provide a sanity check.

\articlefigure[width=.9\textwidth]{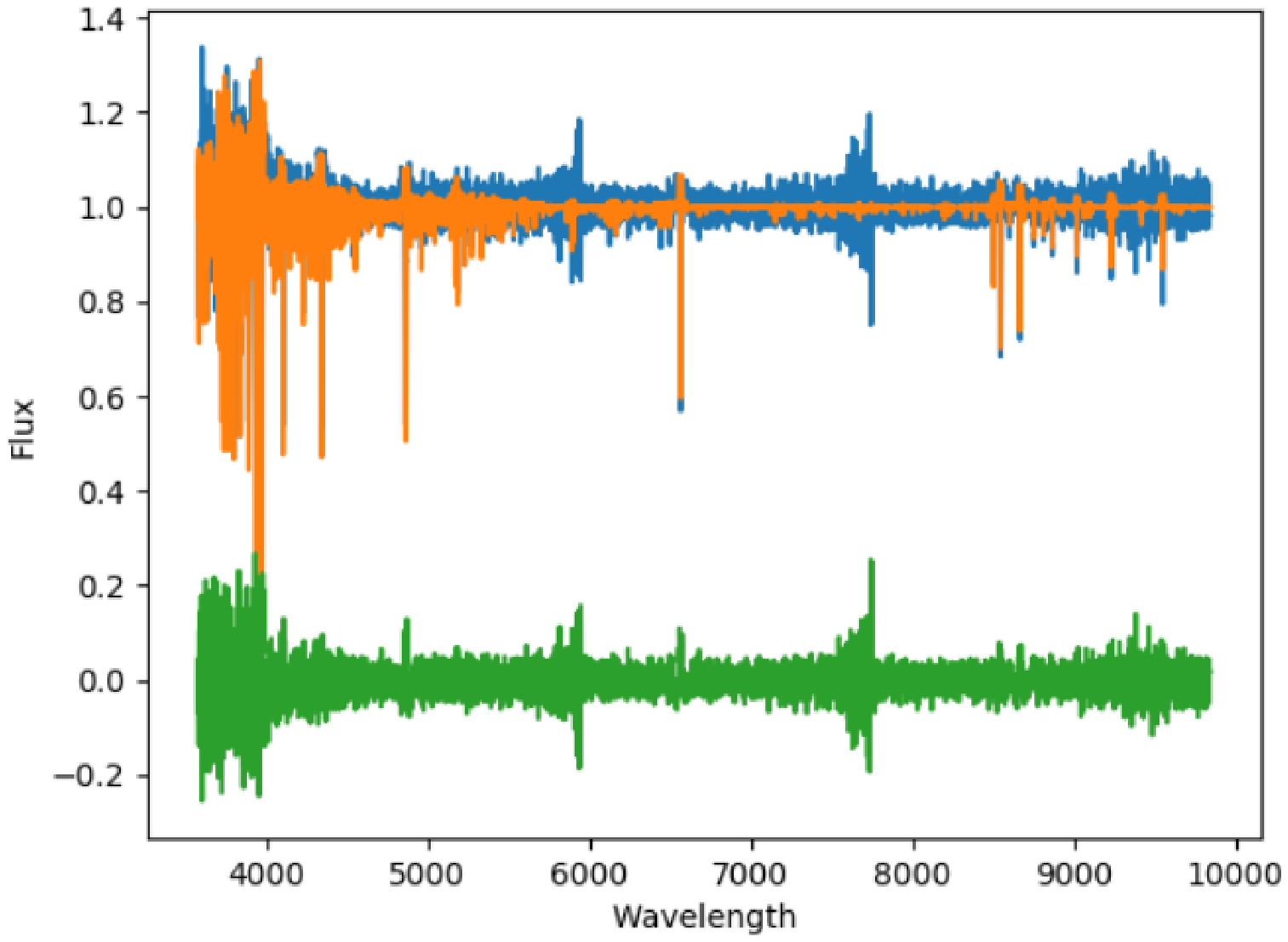}{f2}{Analysis of a continuum-normalized observation of an F-type halo turn-off star. The observed spectrum (blue) is in fact a DESI simulation. The best-fitting model is shown in brown, and the residuals are in green.}

The strategy is to extract the {\it standard's} spectra, remove the instrument response by continuum normalization, and analyze them with synthetic spectra computed from model atmospheres to constrain the atmospheric parameters: effective temperature, surface gravity, and overall metal abundance. Fig. \ref{f2} illustrates this procedure for a simulation of an F-type metal-poor {\it standard}. Our tests suggest this analysis can constrain well effective temperature ($\sim 100$ K) and metallicity ($\sim 0.2$ dex), while surface gravity is more poorly determined ($\sim 0.8$ dex). This is good enough to provide a reliable flux calibration to within a few per cent. 

\section{A test on BOSS data}

Due to the simplifications involved, simulations can sometimes be overly optimistic. For the estimates discussed above, flux losses due to atmospheric differential refraction, telescope pointing and tracking errors, PSF variations across the focal plane and with time, or issues with the {\it standard} stars chosen (e.g. binarity or activity) not being well-represented by classical model atmospheres, are fully ignored. We turn our attention to the analysis of real data from the Baryon Oscillation Spectroscopy Survey (BOSS; Dawson et al. 2013), already mentioned in the preceding sections. 

BOSS was mainly devoted to acquiring galaxy spectra. Nevertheless, there were special plates for cross-survey calibration, overlapping with the SDSS (in particular the Sloan Extension for Galactic Exploration, SEGUE; Yanny et al. 2009) or the Gaia-ESO survey (Gilmore et al. 2012). These plates have large numbers of stars, between 600 and 900, in them, and are ideally suitable to evaluate the quality of the BOSS flux calibration procedure. 

We proceeded as explained in the previous section, fitting the continuum-normalized spectra of F-type stars in the special plates with standard models. Our choice of models was our recent collection (Allende Prieto et al. 2018), based on the latest generation of ATLAS9 Kurucz models (M\'esz\'aros et al. 2012) and spectral synthesis with the ASSET (Koesterke 2009) code. The fittings were carried out with the FERRE\footnote{github.com/callendeprieto/ferre} code. Once we have the best-fitting parameters for each {\it calibrator}, we can then assign theoretical SEDs to them. Accounting for extinction, we can then compare the expected SED with the ones in the SDSS/BOSS data base. 

\articlefigure[width=.9\textwidth]{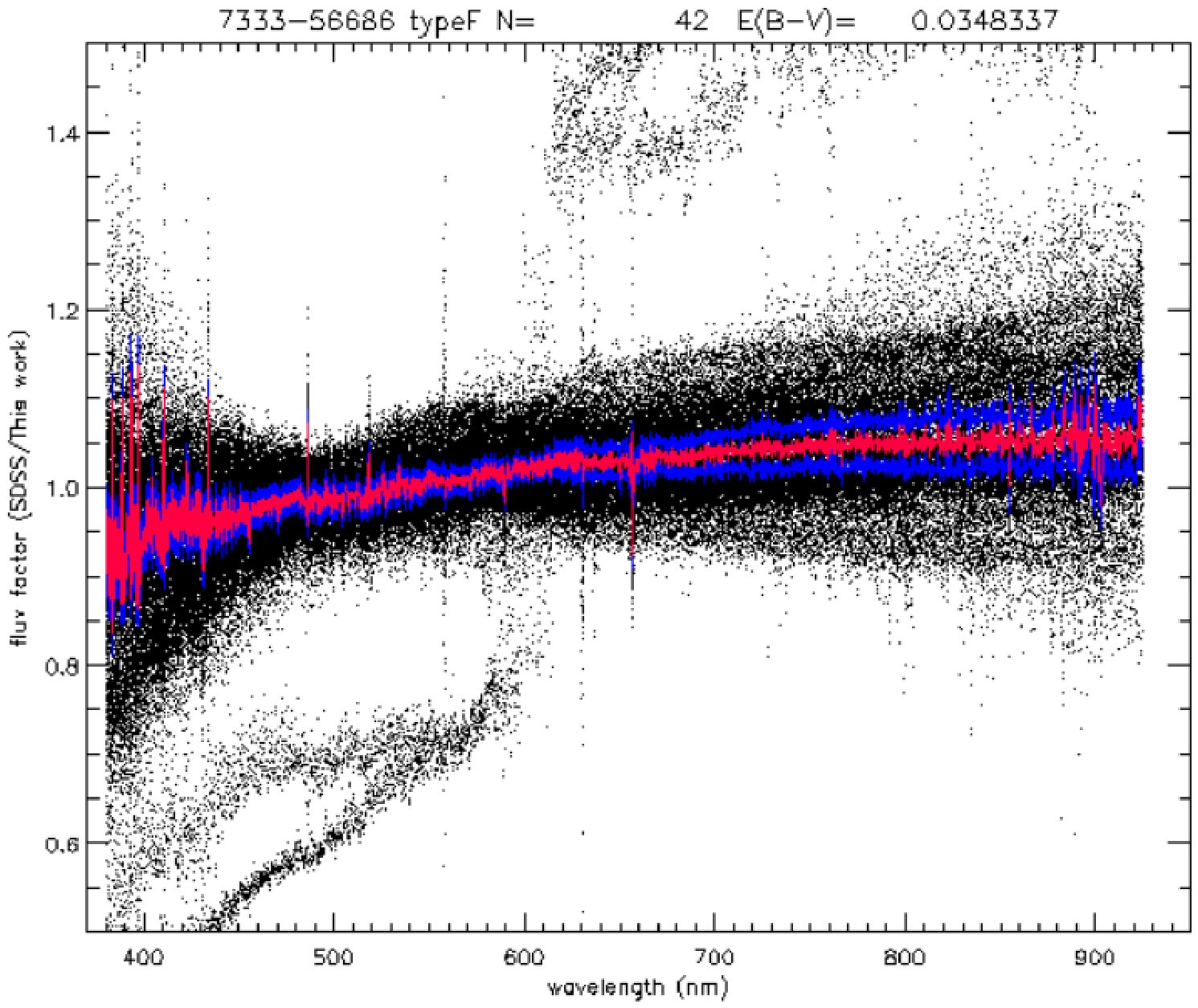}{f3}{Ratio between the fluxes predicted by a model corresponding to the best-fitting atmospheric parameters, derived from continuum-normalized observations, and the actual flux-calibrated BOSS spectra, after normalizing both to have an average flux of unity. The data points corresponding to 42 chosen F-type star calibrators are in black, while the median ratio is in red and the uncertainty of the mean is in blue. The model fluxes have been corrected to include the expected average interstellar extinction in the direction of the field (BOSS plate 7333).}

Fig. \ref{f3} shows the ratio of the BOSS fluxes and those expected from our spectral analysis. The red curve shows the median ratio as a function of wavelength. The median shows mild variations from unity, of about  $\pm 5$ \%. However, the ratios for individual stars can deviate by  20\% or more. 
While the flux calibration of the BOSS observations is, on average, good to a few percent, the same is not true for individual stars. 

The situation for future surveys such as DESI may be better than for BOSS. Gaia astrometry is now available, with an additional flexibility provided by the robotic positioner compared to the aluminum plates used in BOSS. DESI counts on an ADC to minimize chromatic distortions associated with atmospheric refraction. Finally, the DESI spectrographs will be resting on a thermally-controlled room, rather than hanging on the Cassegrain focus of the telescope, as it was the case for SDSS/BOSS, giving them further stability.  

\section{Conclusions}

Stellar spectrophotometry is a valuable source of information on stellar parameters and extinction, but requires extra care to keep instrumental/atmospheric distortions under control. Flux calibration strategies should consider more than one type of calibrator.

Model-based flux calibrations, like those performed using F-type halo turn-off stars in SDSS, are statistically reliable, but at least in the case of the BOSS data can show notable errors, of up to 20\% or more, for individual stars. 

Fiber-fed multiplexed spectrographs pose bigger challenges for flux calibration than more traditional single-target long-slit spectrographs. However, the design of upcoming instruments includes improvements that should result in more accurate spectrophotometry.

%\clearpage % To force this stuff to happen by this point in the text, otherwise these will probably end up after the references.

\acknowledgements Thank you, Ivan: For teaching me physics. For being a model of scientific practice and generosity. For your support. And for the good times, passed and future!

%\bibliography{editor}  % For BibTex

% For non-BibTex:

\end{document}